\documentclass[epj,nopacs]{svjour}
\usepackage{graphics}
\usepackage{graphicx}
\usepackage[figuresright]{rotating}
\usepackage{psfig}
\usepackage{epsfig}
\usepackage{subfigure}

\newcommand{\pt}{p$_\mathrm{T}$}

\begin{document}

\title{Measurement of low mass dielectron continuum in $\sqrt(s_{NN})$=200GeV Au+Au collisions in the PHENIX Experiment at RHIC}
       
\author{Alberica Toia\inst{1} for the PHENIX Collaboration}
\institute{Stony Brook University, NY 11794-3800, USA}
\date{Received: date / Revised version: date}

\abstract{
The first measurement of the dielectron continuum at RHIC energies was performed by the PHENIX experiment for Au+Au collisions at $\sqrt{s_{NN}}$= 200 GeV. 
Mass spectra for different centralities are presented and compared with the expectations from hadron decays.
}

\maketitle

\section{Introduction}
Electromagnetic probes are ideally suited to investigate hot and dense matter produced in high energy heavy ion collisions because they do not undergo strong interactions and thus probe the time evolution of the collision. 
The dielectron continuum is rich in physics. In the low mass region Dalitz decays of light hadrons and direct decays of vector mesons, which might be modified in the medium contribute to the spectrum. 
A number of experiments (E325\cite{Kek}, DLS\cite{Dls}, CERES\cite{Cer} and more recently NA60\cite{Na60_rho}), independent of bombarding energy, have observed an excess of dielectron yield over the hadronic sources by a factor 2-3 for masses between 0.2 and 0.8 GeV/c$^2$, when going from proton to heavy ion induced reactions. 
This enhancement has been interpreted as thermal radiation from pion annihilation in the hot fireball largely mediated by light vector mesons $\rho$, $\omega$ and $\phi$. Among these, the most important contribution arises from the $\rho$ meson, due to its strong coupling to the $\pi\pi$ channel and its short lifetime of only 1.3 fm/c which, in contrast to the longer living $\omega$ (23.4 fm/c) and $\phi$ (44.4 fm/c), makes it very suitable to probe in-medium modification close to the QCD phase boundary. \\
In the intermediate mass region between the $\phi$ and the J/$\Psi$ vector meson the dominant contribution arises from correlated charm production. 
This region has been proposed as an interesting candidate to search for thermal radiation emitted as a blackbody radiation from the QGP, since its contribution could be comparable to that of the charm and would be dominant to higher $p_T$ \cite{rap2}. Earlier measurements \cite{Na50} indicate an excess of yield in nuclear reactions with respect to the rate expected from elementary reactions. More recently NA60 \cite{Na60_therm} has disentangled the prompt dimuons from the pairs originated from open charm semi-leptonic decays and demonstrated the prompt origin of the excess yield.\\
Although correlated $e^{+}e^{-}$ pairs are rare, the 0.24 nb$^{-1}$ collected by PHENIX for Au+Au collisions at $\sqrt{s_{NN}}$= 200 GeV in 2004 provide a significant sample to investigate the dilepton continuum. 
In the following, the analysis of 800M events is presented.

\section{Di-Electron Analysis}
The PHENIX detector design is optimized for electron measurements. It combines an excellent mass resolution (1\,\%) with a powerful particle identification achieved by matching the reconstructed tracks with the information from a Ring Imaging Cherenkov detector (RICH) and an Electromagnetic Calorimeter (EMC). 
Electrons are identified by requiring at least 3 phototubes matched in the RICH and by correlating the energy measured in the EMC and the momentum, parametrized in terms of $\frac{E-p}{p}$.
Pair cuts are applied to avoid sharing of detector hits, tracks which are parallel in the RICH 
are rejected by a cut on the angular difference. Whenever encountering such a pair, the complete event is rejected. \\
Pairs are created by combining all electrons with all positrons in one event. The overwhelming yield of these pairs is unphysical.
A statistical procedure is used to determine this combinatorial background.
Since the PHENIX acceptance is different for like and unlike sign pairs, 
the combinatorial background is computed with a 
mixed-event technique by pairing un-like sign tracks from different events within similar topology (i.e. vertex position and collision centrality).
The like-sign pairs prove that the shape of the background is reproduced with high precision by the mixed event technique, 
since the ratio of the like-sign spectra for real and mixed events deviates from 1 by less than 0.1\,\%. \\

Four different methods have been tested to normalize the background
distributions.
One normalizes the number of mixed events to the number of physical events. 
The second one relies on the mean number of single electron tracks which contribute to uncorrelated pairs
$\langle N_{+-} \rangle
=  \overline{n_+} \cdot \overline{n_-}$.
Both these methods normalize directly or indirectly to the product of the average electron and positron multiplicities, which is equivalent to the assumption that the sources of electron and positron tracks are independent. Unless the probability density is a Poisson distribution an additional correction factor
\begin{equation} 
\zeta = 1+ \frac{\sigma^2 - \langle N \rangle}{\langle N \rangle ^2}
\end{equation}  
is introduced to account for the primary multiplicity distribution.
\begin{figure}[!h]
 \begin{minipage}[l]{0.5\textwidth}
 \epsfig{file=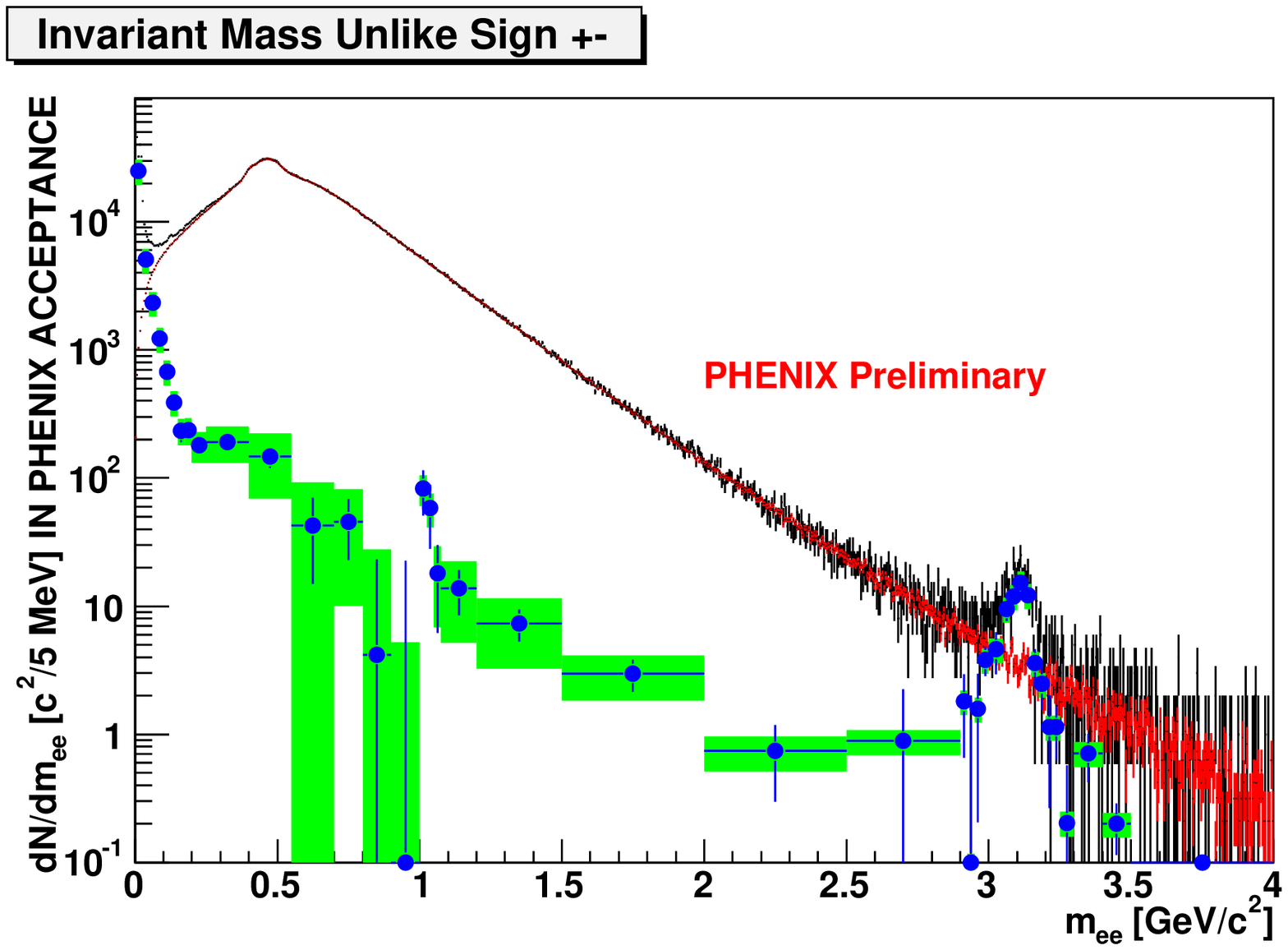,width=\textwidth}
 \end{minipage}
 \hfill
 \begin{minipage}[l]{0.5\textwidth}
 \epsfig{file=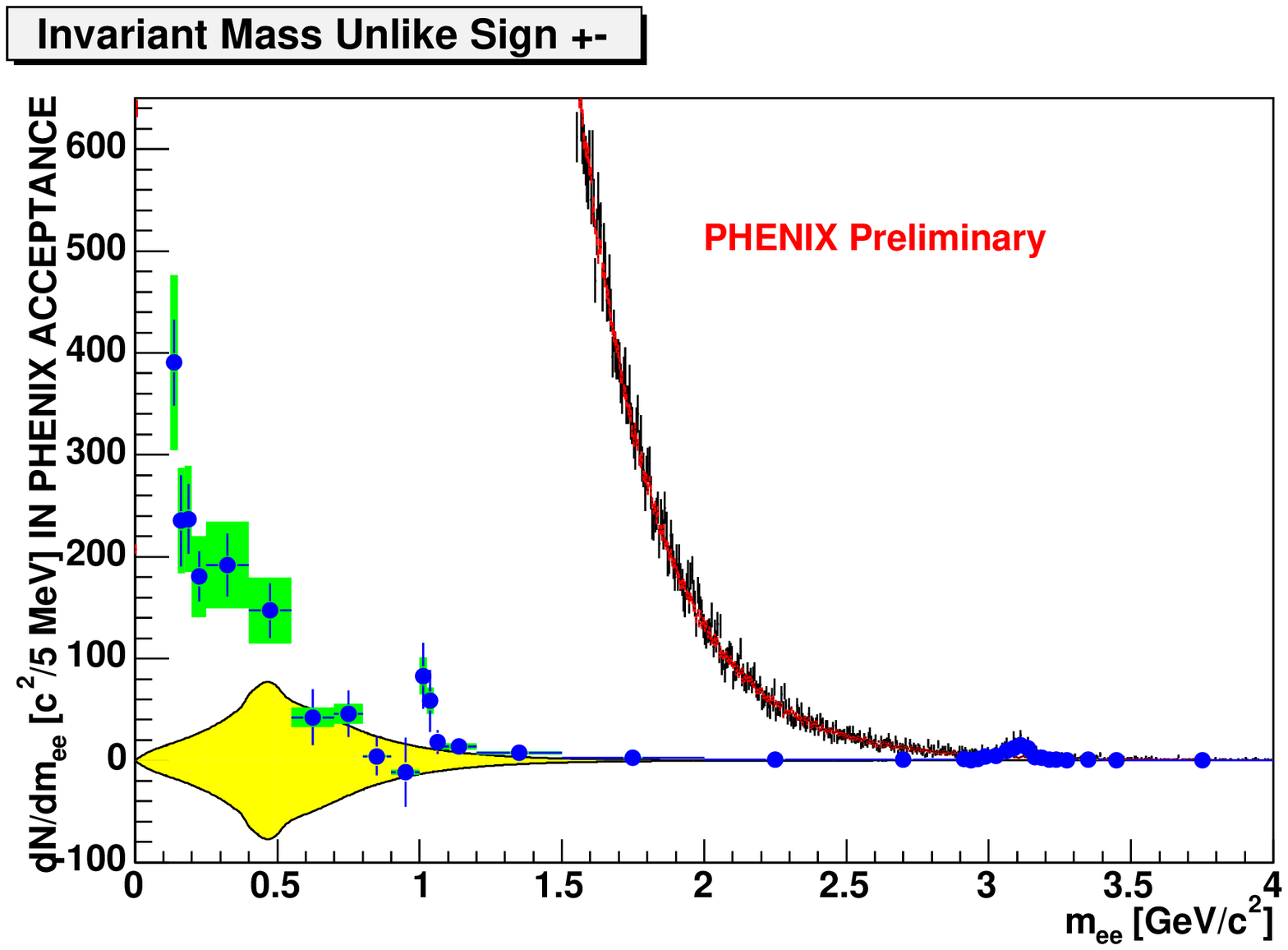,width=\textwidth}
 \end{minipage}
 \caption{Unlike sign mass spectrum: foreground (black), background (red), subtracted spectrum (blue), with systematic error marked in green (all the systematics except for the background) and yellow (systematics of the combinatorial background). Colors in the online version.}
  \label{fig:mass}
\end{figure}
Moreover in heavy ion collisions the probability to produce N pairs depends on other variables which will cause additional correlations in the number of pairs produced in an event. The most relevant are multiplicity correlations due to the finite width of centrality classes, which are characterized by the number of participants $N_{part}$. 
The corrections have been evaluated with a Monte Carlo based on the Glauber model parameterization of $N_{part}$ and $N_{coll}$ and are given by 
\begin{equation} 
\xi = \frac{\langle N_{\pm} \rangle}{\langle N_{+} \rangle \langle N_{-} \rangle }
\end{equation}  
Similar studies have demonstrated that correlations due to the finite size of vertex classes are negligible, considering the low multiplicity of the electron sample.\\  
However, due to the fact that electrons and positrons are always created in pairs, the unlike sign background is the geometric mean of the like sign backgrounds, independent of the primary multiplicity distribution
$\langle N_{+-}\rangle\,=\,2\cdot\sqrt{N_{++}N_{--}}$.
This normalization turns out to be statistically correct and insensitive to multiplicity correlations like the first two. However, due to the presence of a small signal in the like sign yield due to double conversion and double Dalitz decays, the geometrical mean of the like sign yields overestimates the combinatorial background.   
This problem can be repaired by normalizing the background to the foreground in the
like-sign distributions and excluding the low-mass region where a signal is present; since the mixed events produce the proper rate of like and unlike sign pairs, the same factor is then applied to the unlike sign distribution. \\
Any cut applied on a pair can potentially distort the relative rate of like and unlike sign yield. The effect of pair cuts on the background normalization has been studied with a Monte Carlo simulation and survival probabilities for like and unlike sign pairs have been computed as a function of the event topology. Therefore an absolute ($\kappa_{\pm}$) or relative ($\kappa= \frac{\kappa_{\pm}}{ \sqrt{ \kappa_{++}\cdot \kappa_{--}}}$) pair cut bias correction has to be applied in the first two or in the last two cases respectively. \\
Finally all the normalizations agree within 0.5\,\%. 
The empirical normalization of the like-sign pairs was chosen for the final results and 
a systematic uncertainty of 0.25\,\% was assigned. \\
After removal of the combinatorial background, the contribution of photon conversion was removed, cutting on
the orientation angle of the pair in the magnetic field.
Finally the spectra are corrected for efficiency such that the data represent the dielectron yield produced in a collision for which the electron and the positron are both in the detector acceptance.

\section{Results}
Figure \ref{fig:mass} shows the invariant mass spectrum of the measured dielectron pairs, the background, and the subtracted yield with uncertainty: the error bars in the upper panel correspond to all the systematic errors, which are dominated by the background normalization.
The lower panel shows separately the uncertainty arising from electron identification, efficiency, acceptance and run-by-run fluctuations, estimated around 22\,\% represented by the point-to-point green band, and the very small uncertainty on the combinatorial background (0.25\,\%) represented by the yellow band around zero, which becomes the dominant source of the signal uncertainty due to the low signal to background ratio.\\
\begin{figure}[htb]
 \begin{minipage}[l]{0.5\textwidth}
 \epsfig{file=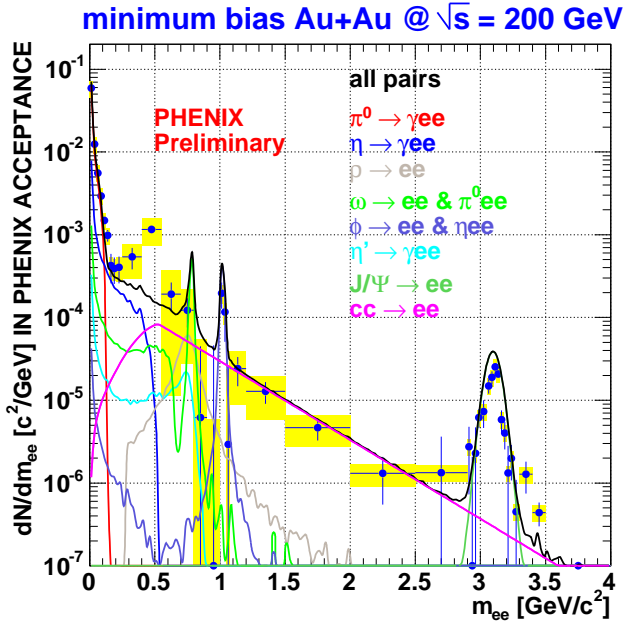, width=\textwidth}
 \end{minipage}
 {\caption{Data compared to a cocktail from all the hadronic sources. The systematic uncertainty of the data is shown in yellow. Colors in the online version.
  \label{fig:cock}}}
\end{figure}
A significant signal is left over the full mass range, corresponding to an integral of $1.8 \cdot 10^5$ pairs, out of those 15,000 above the $\pi^0$ mass. 
The data have been compared to a cocktail of hadron decay sources. 
The pion spectrum used as input is determined by a parametrization of PHENIX charged and neutral
pions. The spectra of the other mesons are determined from the pion
spectrum by $m_T$ scaling \cite{hep0508034}, i.e. using the same Hagedorn
parametrization as for pions and replacing \pt~with $\sqrt{p_T^2 +m_{meson}^2 -m_{\pi^0}^2}$. 
The normalization of the yield relative
to $\pi^0$ is determined by the asymptotic ratio at high \pt~(5\,GeV/c is used here) \cite{Prl94} \cite{eta}.\\
The systematic error, depending on the pion yield and the relative cross section of the other contributions, 
varies from 10\,\% to 25\,\%.\\
An additional source of dielectron pairs, which becomes the dominant continuum contribution for invariant masses above 0.5\,GeV/c$^2$ is the correlated charm production. 
This contribution has been simulated with PYTHIA, scaling the p+p equivalent $c\overline{c}$ cross section of 622$\pm$57$\pm$160 $\mu$barn to the number of minimum bias Au+Au binary collisions \cite{Prl94}. However it is worth to recall that the single electron measurement \cite{ech} shows a suppression at high $p_T$ which increases with centrality. 
The implications for a dielectron invariant mass are not straightforward, since the invariant mass implies knowledge both of the momentum of the two electrons and the opening angle of the pair. 
The charm curve here is meant more as an indication than as a real quantity to compare with.
The systematic error on the charm is therefore meant as an error on the $c\overline{c}$ cross section only, not on the shape.
The calculated electron pairs from charm as well as from the cocktail, have been filtered into the PHENIX acceptance.\\
\begin{figure}[htb]
 \begin{minipage}[l]{0.5\textwidth}
 \epsfig{file=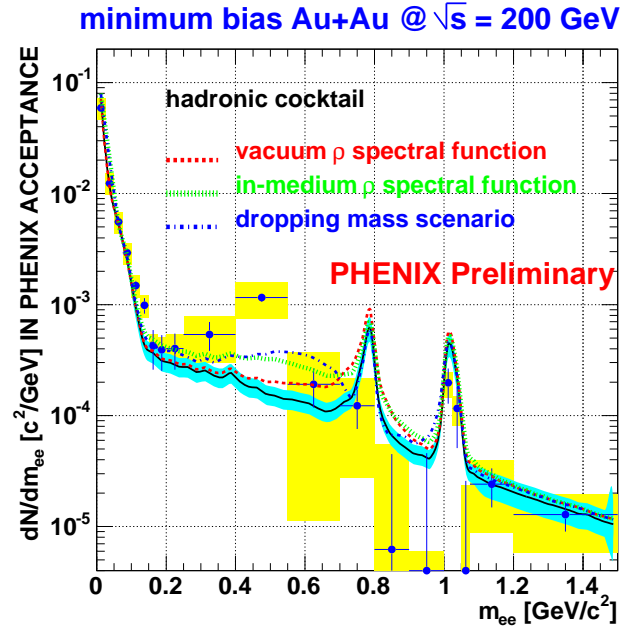, width=\textwidth}
 {\caption{Data (systematic uncertainty in yellow) compared to the cocktail (systematic uncertainty in cian) and theoretical predictions, where a rho spectral function is introduced, in vacuum (red), with a dropping (blue) or melting (green) scenario. Colors in the online version.
\label{fig:rapp}}}
 \end{minipage}
 \hfill
\end{figure}
Figure \ref{fig:cock} shows the data with the total systematic error compared to the cocktail.
The data show a good agreement with the cocktail over the full mass region. The $\omega$ and $\phi$ resonances are not fully reproduced, most likely because of the combined effect of mass resolution and low signal-to-background ratio. 
The data overshoots the cocktail in the region 0.3-0.8\,GeV/c$^2$. The systematic uncertainty however does not allow any strong conlusive statement. \\
In Figure \ref{fig:rapp} the data have also been compared to the theoretical predictions \cite{rap1},\cite{rap2}, where the 
$e^+e^-$ invariant mass spectrum $dN_{ee}/dM$ has been calculated 
using different in-medium $\rho$ spectral
 function and an expanding thermal fireball model. 
Although the systematical uncertainty does not allow any claim of a significant deviation from the known expected sources, it is interesting to observe that the data are even above the predictions that include in-medium modifications in the same mass region where other experiments quantified the same effect.

\section{Centrality dependency}
\begin{figure*}[!ht]
 \begin{minipage}[l]{0.5\textwidth}
 \epsfig{file=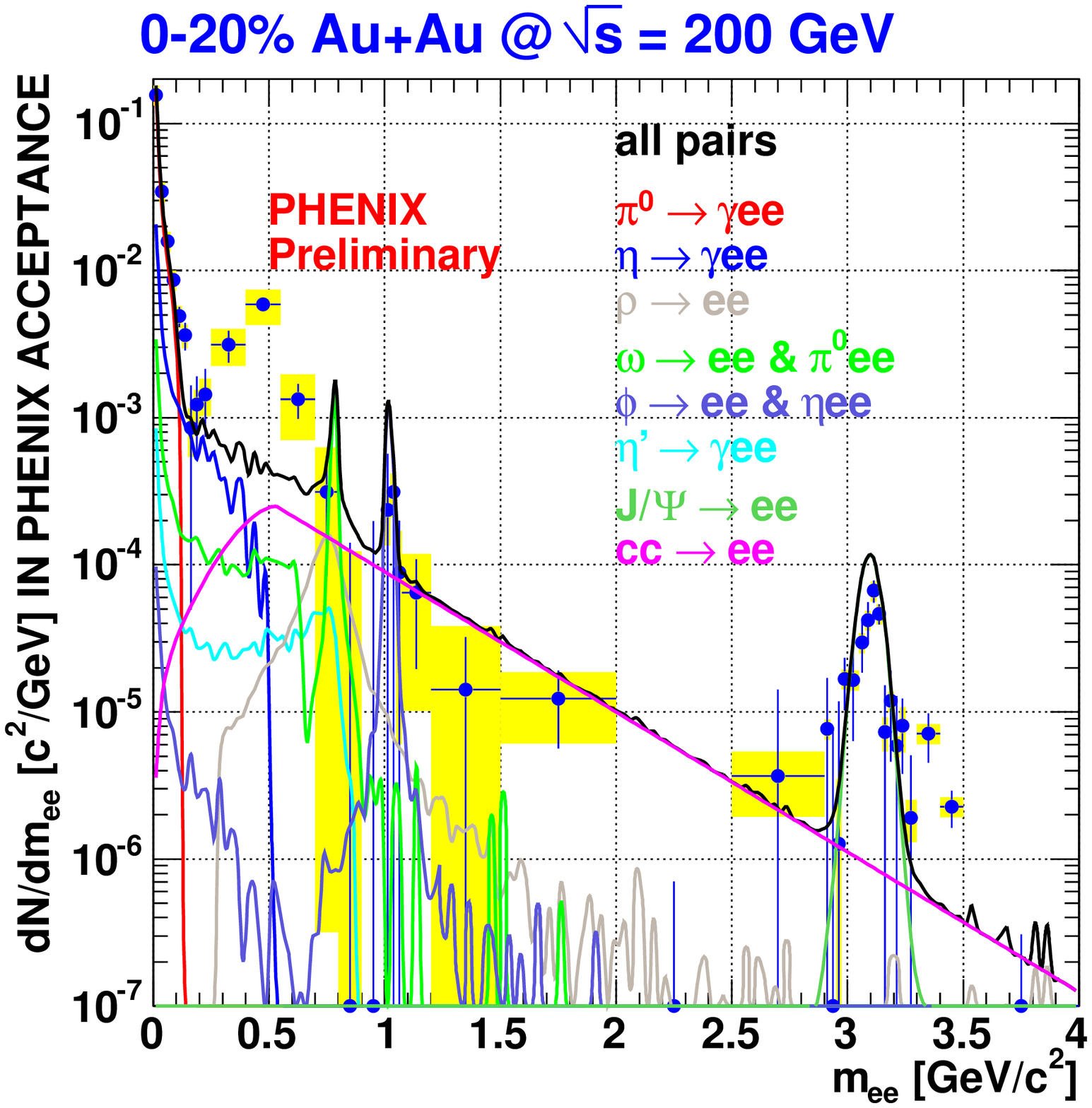,width=\textwidth}
 \end{minipage}
 \hfill
 \begin{minipage}[l]{0.5\textwidth}
 \epsfig{file=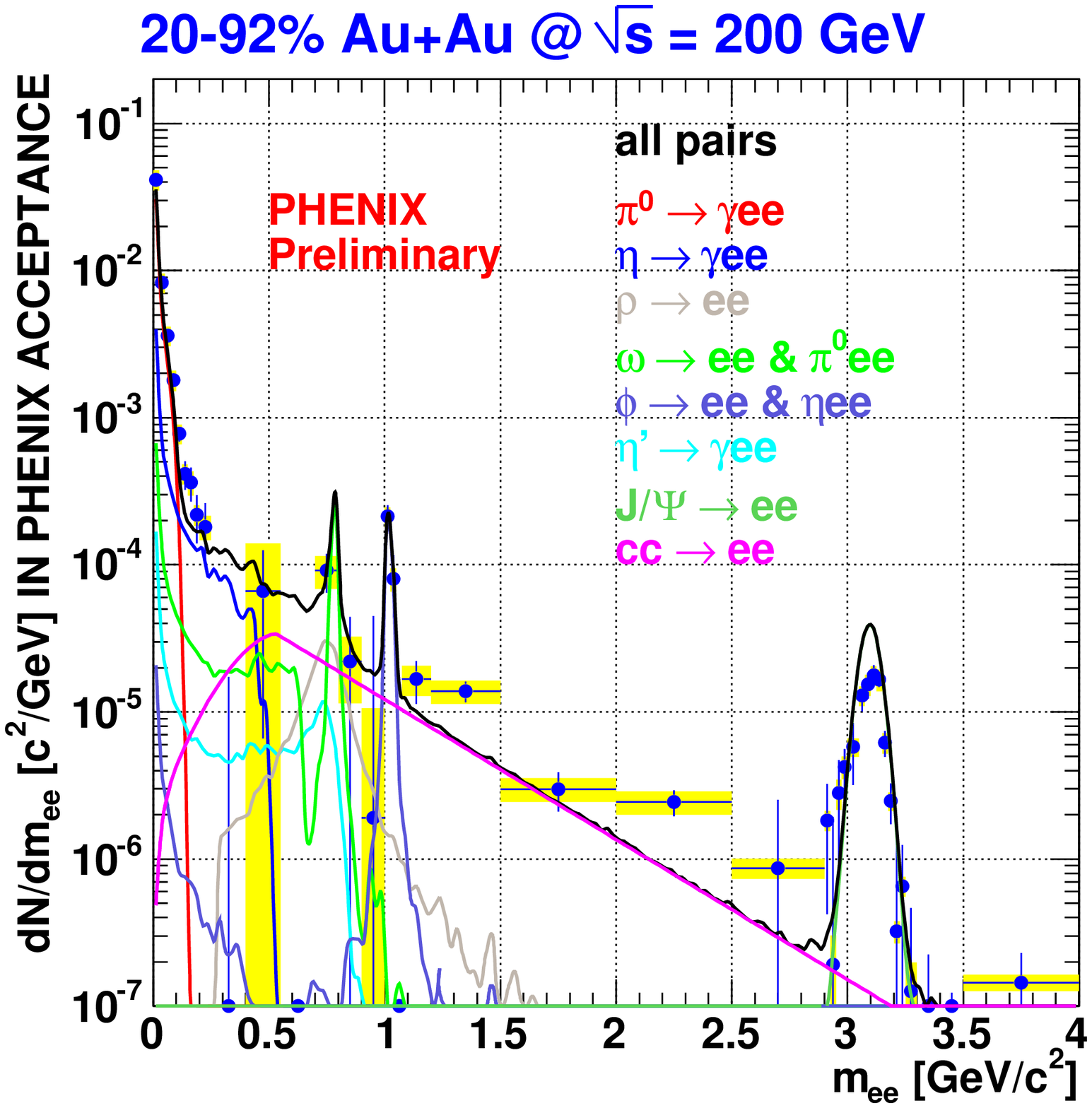,width=\textwidth}
 \end{minipage}
 \caption{Data compared to a cocktail from all the hadronic sources and the charm contribution and ratio for centrality classes 0-20\,\% (right) and 20-92\,\% (left). The systematic uncertainty of the data is shown in yellow. Colors in the online version.}
  \label{fig:cock2}
\end{figure*}
\begin{figure*}[!tb]
 \begin{minipage}[l]{0.5\textwidth}
 \epsfig{file=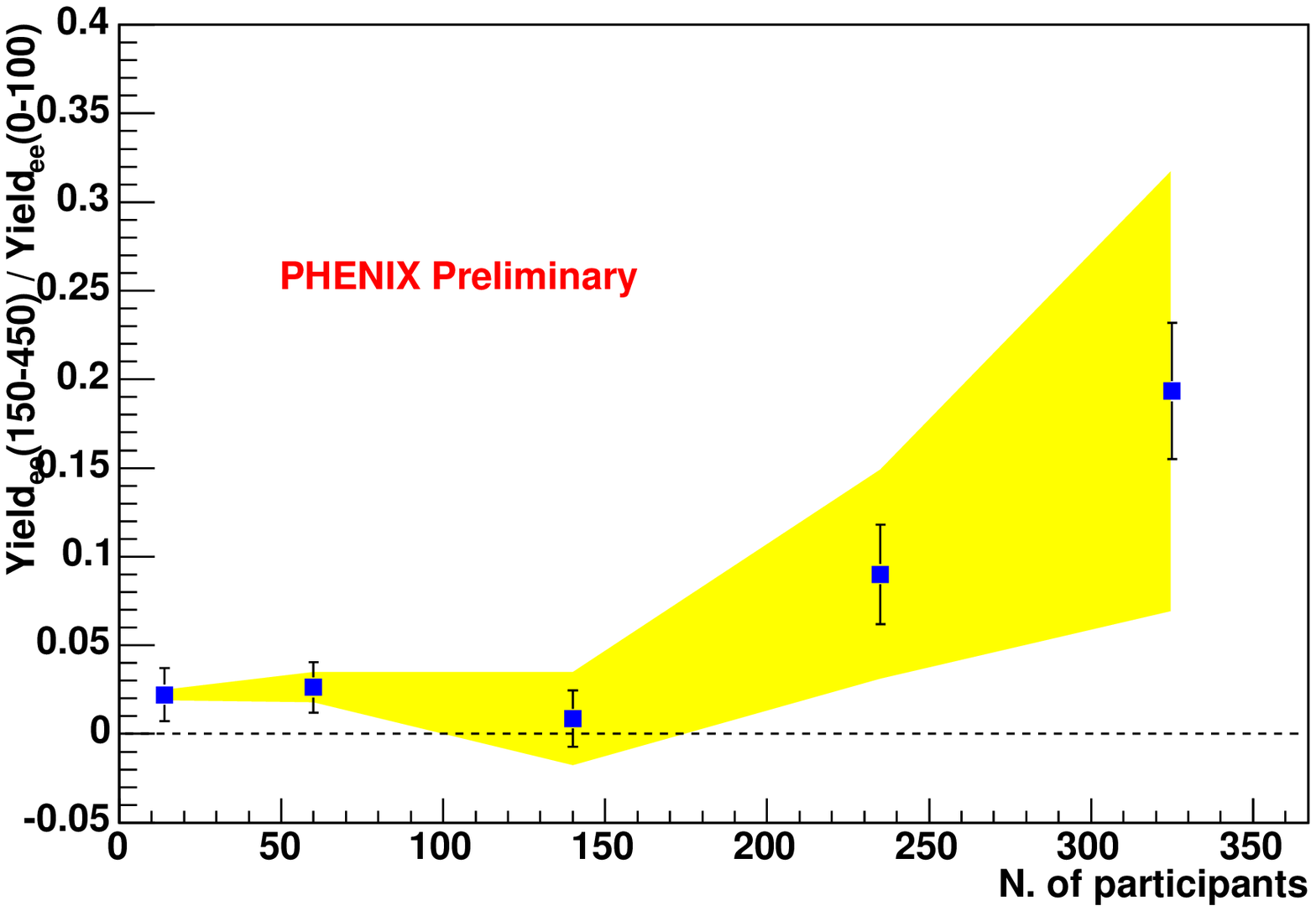,width=\textwidth}
 \end{minipage}
 \hfill
 \begin{minipage}[l]{0.5\textwidth}
 \epsfig{file=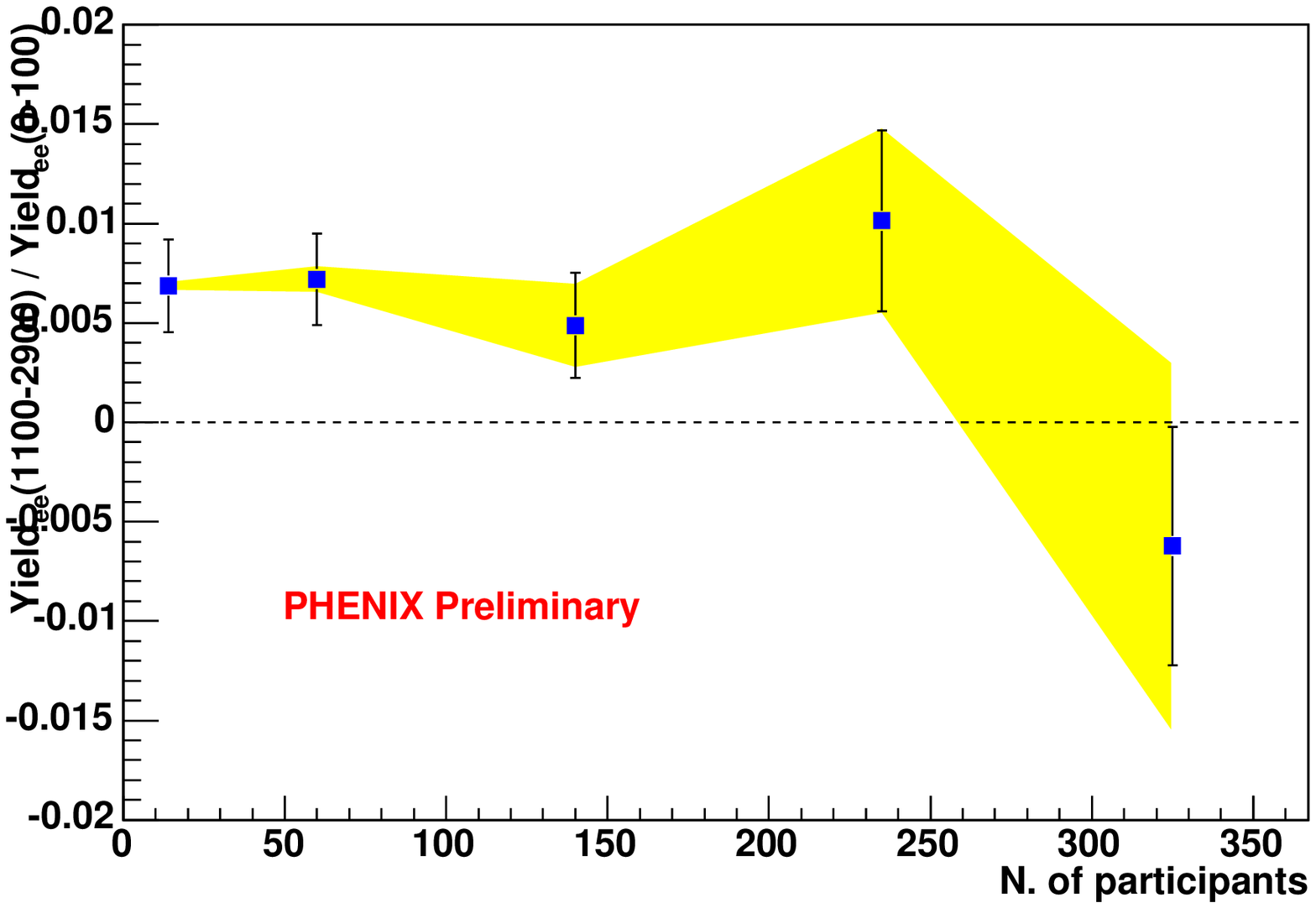,width=\textwidth}
 \end{minipage}
 \caption{Ratio of integral of the mass region 150-450 MeV/c$^2$ (left) and 1.1-2.9 GeV/c$^2$ (right) with respect to the $\pi0$ yield (0-100 MeV/c$^2$). The systematic error, which depends on the combinatorial background in the region 1.1-2.9 GeV/c$^2$ is indicated by the yellow band. Note the different scale in the two plots. Colors in the online version.}
  \label{fig:ratio}
\end{figure*}

Despite the low statistical significance of the dielectron signal, the centrality dependency of the dielectron continuum has been studied. 
The centrality of the collision is determined using the combined information of the Beam-Beam Counter Detectors and the Zero-Degree Calorimeter on particle multiplicity, and translating this information into a centrality notion with the Glauber model.
The event mixing as well as the normalization of the combinatorial background has been performed in each centrality class separately, always using the like-sign spectra as a cross check with an agreement better than 0.1\,\%. Cocktails from hadronic sources have been prepared from the input pion distributions in each centrality class \cite{hep0508034}.\\
Figure \ref{fig:cock2} shows data from two centrality classes: central (0-20\,\%) and semiperipheral (20-92\,\%) compared to the relative cocktails, absolutely normalized. The more central collision show a significant enhancement in the mass region between 0.3-0.8 GeV/c$^2$ which is not present in the more peripheral event sample.
\\
The centrality dependency is also studied via the ratio of yield in different mass regions with respect to the $\pi^0$ yield. 
The production of pions in fact scales with the number of participants, with only a slight increase with centrality.
If an in-medium enhancement of the dielectron continuum yield exists, it could in first order arise from $\pi\pi$ or $q\overline{q}$  annihilation. If so the yield should increase faster than proportional to the number of participants $N_{part}$.
Figure \ref{fig:ratio} shows the ratio of different mass regions to the pion Dalitz region (0-100 MeV/c$^2$) as a function of the number of participants. The left hand plot shows the region between 150-450 MeV/c$^2$ where the increase as a function of number of participants is faster than linear.  
In contrast, charm production is known to follow binary scaling \cite{Prl94}. Therefore the yield in the charm region compared to the pion region should increase as the ratio $\frac{N_{coll}}{N_{part}}$. The right hand plot of figure \ref{fig:ratio} shows the region between 1.1-2.9 GeV/c$^2$, where the rather constant ratio does not reflect the expected behavior.

\section{Conclusions}
Thanks to the high statistics collected in 2004, the first measurement of the dielectron continuum at RHIC energy has been performed by the PHENIX experiment.
The dielectron mass spectrum is higher than the expectations from hadronic sources and shows better agreement with theoretical predictions that include modified spectral function. 
Sampling the data in different centrality classes allows to isolate the excess of dielectrons in the most central collisions, while more peripheral collisions show a good agreement with the hadronic cocktail. 
A comparison of the continuum yield in the mass region 150-450 MeV/c$^2$ to the pion yield shows an interesting dependency of the enhancement as a function of the number of participants.
The signal-to-background ratio however limits the statistical significance of these results. PHENIX is currently developing a new hadron blind detector system to address this issue\cite{Kam}.

\end{document}